\documentclass[twocolumn, tighten]{aastex631}

\newcommand{\source}{J1808}

\shorttitle{Hydrogen-triggered Thermonuclear Bursts from SAX J1808}

\begin{document}

\title{Hydrogen-triggered X-ray Bursts from SAX J1808.4-3658? The Onset of Nuclear Burning}

\author[0000-0003-3872-1703]{Sierra Casten} 
\affiliation{Department of Physics and Astronomy, Michigan State University, 567 Wilson Road, East Lansing, MI 48824, USA}

\author[0000-0001-7681-5845]{Tod E. Strohmayer} 
\affiliation{Astrophysics Science Division and Joint Space-Science Institute, NASA's Goddard Space Flight Center, Greenbelt, MD 20771, USA}

\author[0000-0002-7252-0991]{Peter Bult}
\affiliation{Department of Astronomy, University of Maryland, College Park, MD 20742, USA}
\affiliation{Astrophysics Science Division, NASA Goddard Space Flight Center, Greenbelt, MD 20771, USA}

\begin{abstract}

We present a study of weak, thermonuclear X-ray bursts from the accreting millisecond X-ray pulsar SAX J1808.4--3658. We focus on a burst observed with the \textit{Neutron Star Interior Composition Explorer} on 2019 August 9, and describe a similar burst observed with the {\it Rossi X-ray Timing Explorer} in 2005 June. These bursts occurred soon after outburst onset, $2.9$ and $1.1$ days, after the first indications of fresh accretion.  We measure peak burst bolometric fluxes of $6.98 \pm 0.50 \times 10^{-9}$ and $1.54 \pm 0.10 \times 10^{-8}$ erg cm$^{-2}$ s$^{-1}$, respectively, which are factors of $\approx 30$ and $15$ less than the peak flux of the brightest, helium-powered bursts observed from this source. From spectral modeling we estimate accretion rates and accreted columns at the time of each burst. For the 2019 burst we estimate an accretion rate of $\dot M \approx 1.4\textrm{--}1.6 \times 10^{-10}$ $M_{\odot}$ yr$^{-1}$, and a column in the range  $3.9\textrm{--}5.1 \times 10^7$ g cm$^{-2}$.  For the 2005 event the accretion rate was similar, but the accreted column was half of that estimated for the 2019 burst. The low accretion rates, modest columns, and evidence for a cool neutron star in quiescence, suggest these bursts are triggered by  thermally unstable CNO cycle hydrogen-burning. The post-burst flux level in the 2019 event appears offset from the pre-burst level by an amount consistent with quasi-stable hydrogen-burning due to the temperature-insensitive, hot-CNO cycle, further suggesting hydrogen-burning as the primary fuel source. This provides strong observational evidence for hydrogen-triggered bursts. We discuss our results in the context of previous theoretical modeling.

\end{abstract} 

\keywords{stars: neutron --- X-rays: binaries --- X-rays: bursts ---
  X-rays: individual (SAX J1808.4$-$3658) --- radiation: dynamics}

\newpage

\section{Introduction}
\label{sec:introduction} 

Thermonuclear (Type I) X-ray bursts occur when an accreted layer of matter on a neutron star undergoes a thermonuclear runaway \citep{1975ApJ...195..735H, 2006csxs.book..113S, 2021ASSL..461..209G}. The thin shell thermal instability that triggers these bursts occurs when the energy generation rate due to nuclear burning exceeds the local cooling rate in the accreted shell. 

Depending on the accreted fuel composition there can be a large number of nuclear reactions that contribute to the energy production that powers bursts, however, two reactions are thought to be the primary triggers for the thermal instability.  The first is the highly temperature-sensitive triple-$\alpha$ reaction, which burns three helium nuclei to carbon. The other relevant process is the CNO cycle that burns hydrogen to helium via a series of $(p, \gamma)$ reactions on carbon and nitrogen \citep{1965nmss.book.....F}. Also operating in the cycle are several rate-limiting $\beta^{+}$ decays dependent on slower weak reaction rates. The cycle is closed by the $(p, \alpha)$  reaction that converts $^{15}{\rm N}$ back to $^{12}{\rm C}$. This physics leads to two regimes in which the CNO cycle may operate in this context. Above temperatures of about $8 \times 10^7$ K, the rate of the so-called hot-CNO cycle is limited by the $\beta^{+}$ decays that take $^{15}{\rm O}$ and $^{14}{\rm O}$ to $^{15}{\rm N}$ and $^{14}{\rm N}$, respectively. In this regime the energy generation rate becomes temperature {\it insensitive}, that is, thermally stable. However, for temperatures below $\approx 8 \times 10^{7}$ K the CNO energy generation rate remains temperature sensitive, so that, in principle, a hydrogen-burning thermal instability can operate if the accreting shell is cool enough.  

The implications of these physical processes for the production of X-ray bursts were explored in several early, ``classic'' papers. \cite{1981ApJ...247..267F} were among the first to delineate the possible bursting regimes based on the mass accretion rate, but also see \cite{1978ApJ...225L.123J} and \cite{1979ApJ...233..327T}.  They identified three bursting regimes with decreasing mass accretion rate.  At higher accretion rates a helium shell grows via accretion and thermally stable CNO cycle burning that converts some of the accreted hydrogen to helium. The accretion rate is high enough that the base of the shell reaches ignition conditions before all the hydrogen can be burned to helium. The triple-$\alpha$ instability is initiated in a shell with a significant hydrogen abundance, leading to so-called mixed H/He bursts. The presence of hydrogen allows for the more complex set of nuclear reactions known as the rapid-proton capture process to occur, which can enhance and delay the energy release, leading to longer duration bursts  \citep{2001PhRvL..86.3471S}. The well-studied ``clocked'' bursts from GS 1826$-$238 are prototypical of this type \citep{1999ApJ...514L..27U, 2004ApJ...601..466G, 2007ApJ...671L.141H, 2012ApJ...749...69Z}. For lower accretion rates there will be a rate such that, prior to ignition, the CNO burning has just had sufficient time to burn all the hydrogen in the accreting shell. At this critical accretion rate the helium burning is initiated in a pure helium shell.  These ``pure helium'' bursts are characterized by a rapid, intense energy release, are typically of shorter duration than the mixed H/He bursts, and often reach the Eddington limit, as exhibited by photospheric radius expansion (PRE). The bright PRE bursts observed from the accreting millisecond X-ray pulsar (AMXP) SAX J1808.4$-$3658 are examples of such bursts \citep{2006ApJ...652..559G, 2013A&A...553A..83I, 2019ApJ...885L...1B}. At low accretion rates, temperatures in the accumulating shell may be low enough that steady CNO hydrogen-burning essentially switches off, but can proceed in the unstable, temperature-sensitive regime. This can, in principle, lead to unstable ignition of the hydrogen in the accumulating shell. Two possible paths have been discussed in this case. First, ignition of hydrogen raises the temperature in the shell, and if the column depth is large enough the heated shell may cross the helium instability curve, producing a prompt, mixed H/He burst. Second, if the hydrogen ignition depth is too shallow to cross the helium instability curve, then fuel will continue to accumulate until helium ignition can occur. At these low accretion rates it is also likely that gravitational sedimentation of heavier elements relative to hydrogen and helium will play a role in setting the conditions for unstable burning \citep{2007ApJ...654.1022P}.  

However, observational evidence for this hydrogen ignition regime is limited, as there have been to our knowledge few published reports of X-ray bursts that can be clearly attributed to unstable hydrogen shell flashes. In one such case, \cite{2007A&A...465..559B} reported on the first observations of triple, short recurrence time (SRT) bursts from the high inclination, eclipsing source EXO 0748$-$676.   They suggested that the initial bursts of singles, pairs or triples (they call these the long waiting time, LWT, bursts), could be attributed to either helium-triggered, mixed H/He bursts at moderate accretion rates ($10\%$  of Eddington), or perhaps hydrogen-triggered bursts at lower accretion rates ($1\%$ of Eddington).  Because the LWT bursts appeared somewhat under-luminous compared with mixed H/He bursts in 1-d {\it Kepler} models and the well-known example of such bursts from GS 1826$-$238 they suggested that this might be explained by the latter, hydrogen-triggered mechanism. 

They further suggested that the SRT events, with waiting times close to 10-12 minutes, were likely caused by the re-ignition of unburned fuel, but they did not have a detailed explanation of how this occurs.  More recently, \cite{2017ApJ...842..113K} have outlined a theoretical mechanism to account for SRT bursts.  Using detailed, 1-d {\it Kepler} hydrodynamic simulations they showed that such events can be produced by opacity-driven convective mixing that transports fresh fuel to the ignition depth, and they also argued that this mechanism can produce simulated burst events that are ``strikingly similar'' (in their words) to the SRT bursts seen from EXO 0748$-$676.  If this mechanism is indeed at work, then it would further argue for the higher accretion rate ($10\%$ of Eddington), helium-triggered scenario in EXO 0748$-$676, as warmer envelopes, naturally produced by higher accretion rates, were required to produce the SRT events in their burst simulations.  Moreover, they also showed that the fraction of fuel burned in the LWT events dropped as the envelope became hotter, and this relatively low fuel burning fraction could also naturally explain the apparently under luminous LWT bursts noted by \cite{2007A&A...465..559B}.   Thus, while \cite{2007A&A...465..559B} suggest that a hydrogen-triggered mechanism is possible for the LWT bursts from EXO 0748$-$676, we would characterize the current, overall evidence in support of that conclusion as tentative, particularly given the remaining uncertainties in the distance and anisotropy factors for this source.  Indeed, in support of this we note that in their recent review of the field, \cite{2021ASSL..461..209G} also comment that, ``No observations matching case I or case II bursting have been identified.'' Here, cases I and II refer to the two hydrogen ignition paths at low accretion rates that we sketched above.

In this paper we present a study of an apparently rarer class of weak X-ray bursts observed from SAX J1808.4$-$3658 (hereafter, J1808) that we argue show the hallmarks of being associated with the hydrogen ignition regime.  This object was the first AMXP discovered \citep{1998Natur.394..344W, 1998Natur.394..346C}, and hosts a neutron star in a 2.1 hr orbit with a low-mass brown dwarf \citep{2001ApJ...557..292B}. Its distance has been estimated at $3.5 \pm 0.1$ kpc \citep{2006ApJ...652..559G}, and it is likely that the donor provides a hydrogen-rich mix of matter to the neutron star during outbursts \citep{2006ApJ...652..559G, 2019MNRAS.490.2228G}. To date, J1808 has been observed extensively during ten outbursts. While it is not our intention here to provide a broad observational overview of the source--for the purposes of this paper we focus on issues relevant to its thermonuclear bursting behavior--readers can find elsewhere some recent studies on coherent pulse timing \citep{2017MNRAS.471..463S, 2020ApJ...898...38B, 2022arXiv221209778I}, X-ray spectral properties \citep{ 2019MNRAS.483..767D}, and aperiodic timing behavior \citep{2015ApJ...806...90B, 2022MNRAS.tmp.3527S}. 

Observations of J1808 have revealed two types of thermonuclear bursts that show dramatically different peak fluxes and fluences. The bright PRE bursts (mentioned above) show significantly higher total energy release and peak X-ray flux. The less frequently observed weak bursts produce much less energy and show peak fluxes about a factor of 25 less than the bright events, as such they are not Eddington-limited. When these weak bursts have been observed, they appear to be confined to earlier portions of the outbursts and occurred before the bright bursts were seen. This suggests there may be a window of occurrence for these bursts associated with the initial onset of accretion after a period of quiescence. This is particularly intriguing in the context of J1808 because it is known that the neutron star cools dramatically in quiescence \citep{2009ApJ...691.1035H}, and the unstable hydrogen-burning regime requires cooler temperatures in the accumulating layer. There has been more observational and theoretical research exploring the nature of the bright bursts than the weak class.

Here we present a detailed study of one of these weak bursts that was observed with the {\it Neutron Star Interior Composition Explorer} ({\it NICER}) during the recent, 2019 August outburst from J1808. We also provide a briefer description of a similar burst observed with the {\it Rossi X-ray Timing Explorer} {\it (RXTE)} in 2005 June. The paper is organized as follows.  In \S 2 we introduce the {\it NICER} data and present light curves focusing on the initial part of the 2019 outburst, showing a single weak burst.  We also present a spectral study of the persistent and burst emission (for the weak burst) in order to understand its energetics and to constrain the mass accretion rate and the likely accreted mass column at the time of its ignition.  We present a discussion in \S 3 of a likely physical scenario that results in the weak burst, arguing that the initial accretion onto a cool neutron star at the onset of the outburst naturally places the accumulating layer in the thermally unstable regime for CNO hydrogen ignition. Here, we also describe the 2005 June {\it RXTE} event, and we also report a brief summary of {\it NuSTAR} observations that began on 2019 August 10 and in which several brighter bursts were detected.  We conclude in \S 4 with a summary, a brief discussion of relevant uncertainties and other possible interpretations, and the outlook for future efforts.  

\section{NICER Observations of J1808}

\begin{figure}[htbp]
\includegraphics[width=\linewidth]{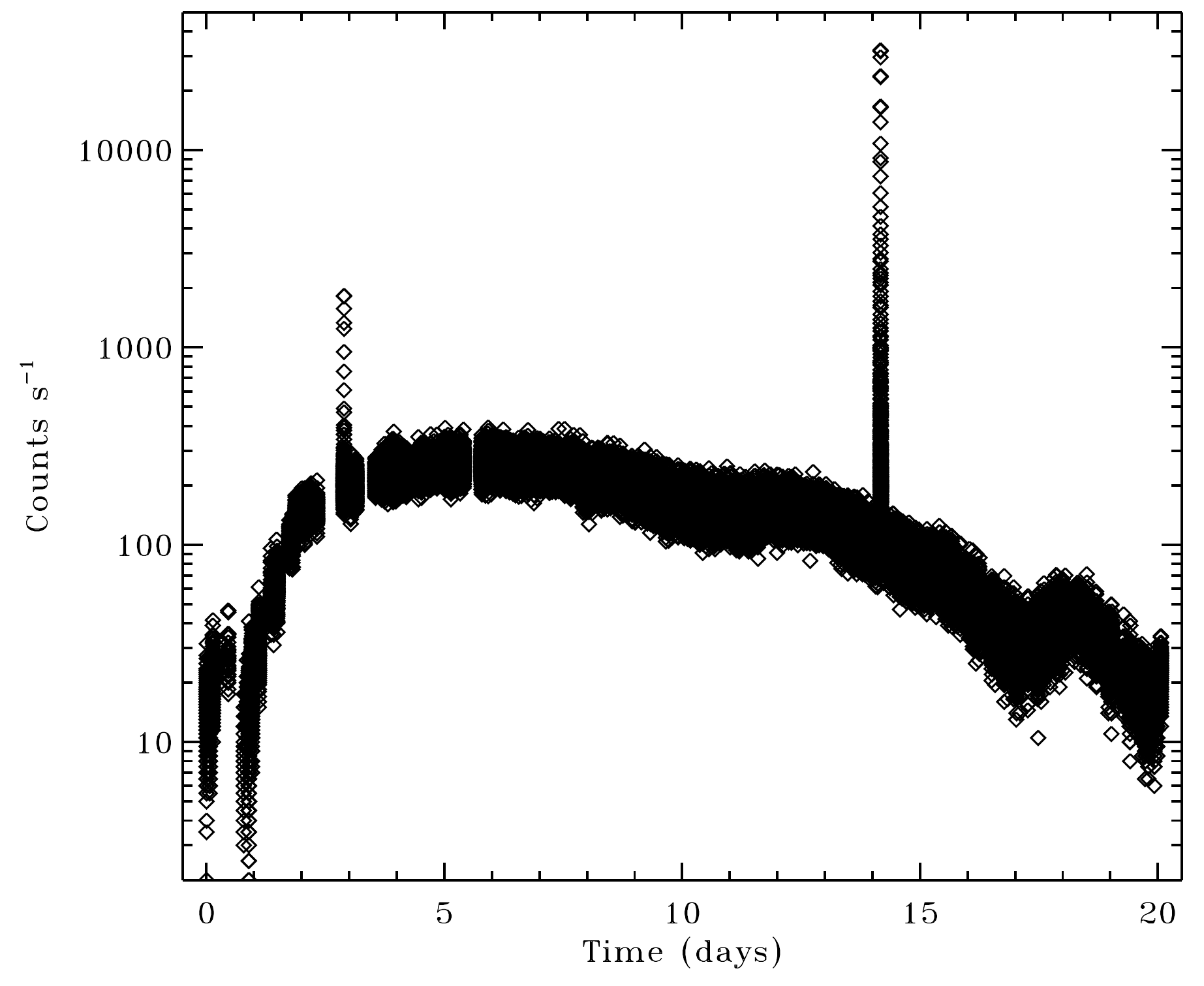}
\caption{\label{fig:totallightcur} Light curve of the 2019 July--August outburst of \source{} in the 0.4--7 keV band observed with {\it NICER}. Count rates were computed in 2 s intervals. Note the logarithmic scale. The first, weaker burst is evident near day 3. The bright burst near day 14 was reported on by \cite{2019ApJ...885L...1B}. }
\end{figure}

In late July 2019, it was reported that the optical flux from J1808 had increased, perhaps presaging a new X-ray outburst \citep{2019ATel12964....1R, 2020MNRAS.498.3429G}. This initiated an extensive monitoring campaign with {\it NICER}, which began on August 1, 2019 \citep{2020ApJ...898...38B}. {\it NICER} is an X-ray observatory that operates on the International Space Station (ISS). It observes across the 0.2--12 keV X-ray band and provides low-background, high-throughput ($\approx 1900$ cm$^{2}$ at 1.5 keV), and high time resolution capabilities \citep{2012SPIE.8443E..13G}.  The data obtained prior to the onset of the outburst, and up to and including the first observed X-ray burst are organized under observation IDs (OBSIDs), 205026010$mm$, and 25840101$nn$, where $mm$ and $nn$ run from 03-10 and 01-02, respectively. We used the standard screening criteria and {\it NICERDAS} version 8 to produce cleaned event lists.  This means we retained only those epochs during which the pointing offset was $< 54''$, the Earth elevation angle was $ > 15^{\circ}$, the elevation angle with respect to the bright Earth limb was $> 30^{\circ}$, and the instrument was not in the South Atlantic Anomaly (SAA). We used HEASOFT Version 6.29c to produce the light curves and spectra for the analyses reported here. The initial observations of the campaign did not reveal evidence of \source{} in X-ray outburst. The first indication that an accretion-driven flux was present occurred on August 6, 2019 at approximately 21:59 TT \citep{2019ATel13001....1B}.  Figure ~\ref{fig:totallightcur} shows the light curve (0.4--7 keV) of the outburst over approximately 20 days from the observed onset of significant X-ray activity.  Time zero in the plot refers to the time of outburst onset, 58701.91597 MJD (TT). The two detected X-ray bursts are evident as ``spikes'' in the count rate near days 3 and 14, respectively. The much brighter second burst (near day 14) was reported on by \cite{2019ApJ...885L...1B}. Here we focus on a study of the much weaker first burst, which occurred at 58704.80764 MJD (TT), and is present in OBSID 2584010102. 

\subsection{Persistent Spectrum, Fluence and Accreted Mass}

To explore the weak burst energetics and ignition conditions we aim to constrain the total accreted mass from the beginning of the outburst up to the onset of the first burst. To do this we model the spectrum of the persistent emission to determine its flux and then integrate from the outburst onset to just prior to the burst. This integral provides an estimate of the energy fluence produced via accretion, which can then be converted to an accreted mass using standard assumptions for the  accretion luminosity produced by the release of gravitational potential energy of the accreted matter. 

In practice we find that the shape of the persistent spectrum gradually changes during this portion of the outburst, with the spectrum showing a modest hardening over time. We therefore measure the flux at a few intervals along the outburst rise, and use these measurements to estimate the flux per unit {\it NICER} count rate. We then use simple linear interpolation and the trapezoidal rule to integrate the flux from outburst onset to the first burst to estimate the energy fluence.  

\begin{table*}[t]
\begin{center}
\caption{Spectral Model Parameters for SAX J1808: Persistent Emission}
\label{tab:persistent emission}
\begin{tabular}{lrrr}
\tableline\tableline
Parameter & 2584010102 (Pre-burst) & 2050260110 & 2050260109 \\
\tableline\tableline
$n_H$ ($10^{22}$ cm$^{-2}$) & $0.131 \pm 0.016$ & $0.160 \pm 0.025$ & $0.177 \pm 0.028$ \\
$kT_{in}$ (diskbb, keV) & $0.849 \pm 0.026$ & $0.679 \pm 0.024$ & $0.528 \pm 0.017$ \\
Norm (diskbb) & $27.33 \pm 5.36$ & $25.19 \pm 6.46$ & $10.82 \pm 1.79$ \\
$kT$ (bbodyrad, keV) & $2.03 \pm 0.15$ & $1.67 \pm 0.15$ & \nodata \\
Norm (bbodyrad, keV) & $0.98 \pm 0.47$ & $0.604 \pm 0.336$ & \nodata  \\
Index (power) & $1.96 \pm 0.37$ & $2.50 \pm 1.05$ & $2.01 \pm 0.23$ \\
Norm (power) & $0.0266 \pm 0.0137$ & $4.64\times 10^{-3}$ & $4.23 \pm 1.6 \times 10^{-3}$ \\
E (gauss, keV) & $0.990 \pm 0.009$ & $0.935 \pm 0.014$ & \nodata \\
$\sigma_E$ (gauss, keV) & $0.015$ & $0.015$ & \nodata \\
Norm (gauss) & $8.44 \pm 2.03 \times 10^{-4}$ & $3.51\pm 1.00 \times 10^{-4}$ & \nodata \\
$f_{0.1-20}$ (erg cm$^{-2}$ s$^{-1}$) & $7.06 \pm 0.23 \times 10^{-10}$ & $2.08\pm 0.40 \times 10^{-10}$ & $5.35 \pm 0.81 \times 10^{-11}$ \\
$f_{bol}$ (erg cm$^{-2}$ s$^{-1}$) & $7.9 \times 10^{-10}$ & $2.5\times 10^{-10}$ & $6.4 \times 10^{-11}$  \\
$\chi^2$ (dof) & 117.9 (112) & $98.9$ (97) & $94.5 (106)$   \\
Rate ($s^{-1}$, 0.5-10 keV) & $181.9 \pm 0.4$ & $58.70 \pm 0.25$ & $13.46 \pm 0.07$ \\
Epoch (d) & 2.89 & 1.45 & 0.59 \\
Exposure (s) & 740 & 927 & 2807 \\
\tableline\tableline
\end{tabular}
\end{center}
\tablecomments{Parameter uncertainties are estimated as $1\sigma$ values. For OBSID 2050260109 the bbodyrad and gauss line components were not required in the fit. The \nodata symbols in this case indicate that these parameters were not included in the fit. For additional context, the ``Epoch'' specifies the center time of the interval in which the spectra were extracted, and the value refers to the time axis of Figure 1. }
\end{table*}

\begin{figure}[htbp]
\includegraphics[width=\linewidth]{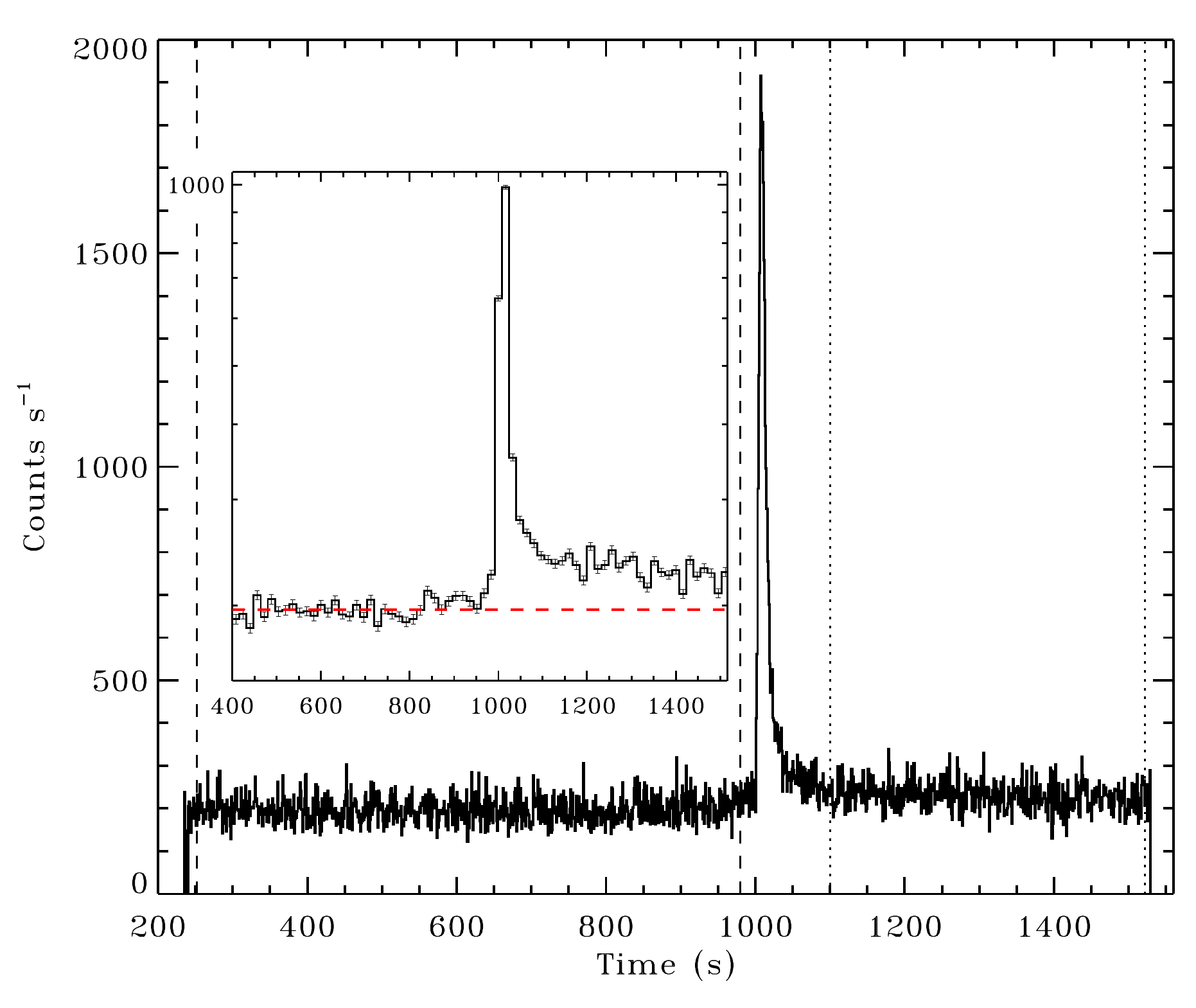}
\caption{\label{fig:lightcurburst} Light curve of the first, weak burst from \source{} in the 0.4 - 7 keV band observed with {\it NICER}. Main panel: The count rates were computed in 1 s intervals, and the vertical dashed and dotted lines denote the intervals used to extract spectra for the pre- and post-burst spectral modeling, respectively. Inset panel: The same data are used, but the time bins are 16 s, and the logarithmic scale highlights the offset in count rate between the pre- and post-burst emission. The dashed red line is a constant value fit to the pre-burst level, and is meant as a guide to the eye. }
\end{figure}

\begin{figure}[htbp]
\includegraphics[width=\linewidth]{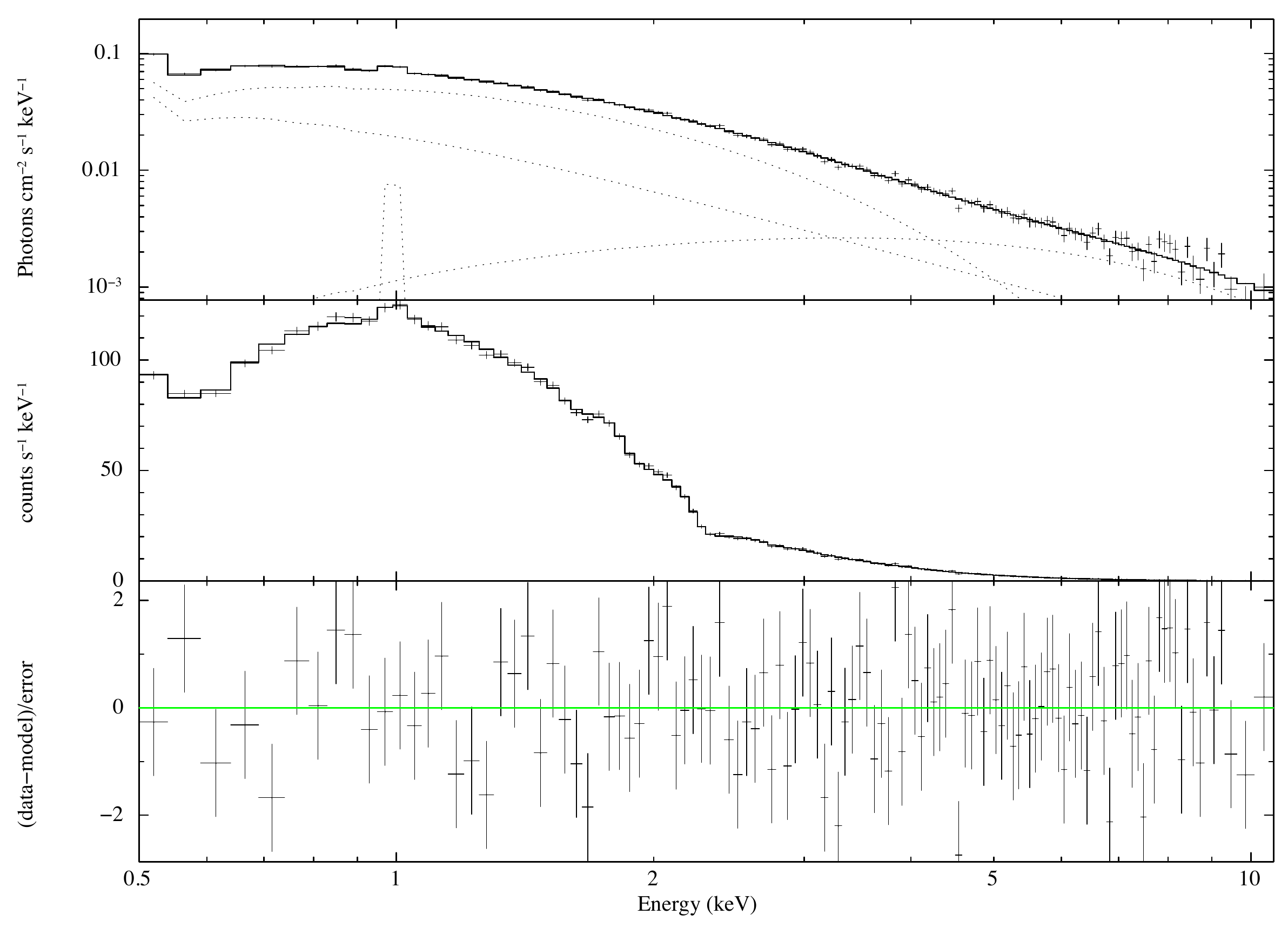}
\caption{\label{fig:energyspec} Energy spectrum of the pre-burst interval, modeled with a phenomenological model similar to that employed by \cite{2009MNRAS.396L..51P}, which includes diskbb, bbodyrad and powerlaw components, in addition to the line at 1 keV. See the text in \S 2.1 for further details.}
\end{figure}

The light curve in Figure \ref{fig:lightcurburst} shows a close-up of the epoch around the first burst. We extracted a spectrum prior to the burst, the ``pre-burst'' interval (marked by the vertical dashed lines in Figure ~\ref{fig:lightcurburst}) and modeled its spectrum using XSPEC version 12.12.1 \citep{1996ASPC..101...17A}.  We produced response files with {\it NICERDAS} version 8, and we used the 3C50 background model, {\it nibackgen3c50} \citep{2022AJ....163..130R}, to produce a background spectrum appropriate for spectral modeling within XSPEC. We employed a phenomenological model similar to that discussed by \cite{2009MNRAS.396L..51P}, that includes thermal disk, power-law, and blackbody continuum components. In addition, and similarly to \cite{2019ApJ...885L...1B}, we find evidence for narrow-line emission near 1 keV, and we include a gaussian component to model this. In XSPEC notation the model has the form, {\it phabs*(diskbb + powerlaw + bbodyrad + gaussian)}, where {\it phabs} represents the line of sight photoelectric absorption model parameterized by the column density of neutral hydrogen, $n_H$. This absorption model uses cross sections from \cite{1996ApJ...465..487V} and the chemical abundances from \cite{1989GeCoA..53..197A}. We fit this model across the 0.5 $-$ 10 keV bandpass and find that it provides an excellent fit, with a minimum $\chi^2 = 117.9$ for 112 degrees of freedom. The best-fitting model parameters are given in Table~\ref{tab:persistent emission}, and Figure ~\ref{fig:energyspec} shows the unfolded photon spectrum (top), the observed count-rate spectrum and best-fitting model (middle), and the fit residuals in units of standard deviations (bottom). This model gives an unabsorbed flux (0.1 $-$ 20 keV) of $7.06 \pm 0.23 \times 10^{-10}$ erg cm$^{-2}$ s$^{-1}$. If we extend the bandpass to estimate a bolometric flux we find a value (0.1 $-$ 100 keV) of $7.9\times 10^{-10}$ erg cm$^{-2}$ s$^{-1}$.  The average count rate in the fitted energy band (0.5 $-$ 10 keV) is $181.9 \pm 0.5$ s$^{-1}$, so we estimate a flux per NICER count rate (0.5 $-$ 10 keV) of $4.34 \times 10^{-12}$ erg cm$^{-2}$ s$^{-1}$ $({\rm counts} \;  {\rm s}^{-1})^{-1}$ for this interval.

We extracted spectra from two other OBSIDs along the outburst rise, 2050260109 and 2050260110, and analyzed these spectra in the same manner as for the pre-burst interval just described. Results of these spectral fits are also reported in Table 1.  For these intervals we estimate flux per NICER count rate values of $2.85 \times 10^{-12}$ and $3.61 \times 10^{-12}$ erg cm$^{-2}$ s$^{-1}$ $({\rm counts} \;  {\rm s}^{-1})^{-1}$, respectively. 

For completeness we make a few additional comments regarding the 1 keV line component included in the spectral model.  For the pre-burst interval (OBSID 2584010102), removing the gaussian line results in an increase in $\chi^2$ of 31.3, and the ratio of the line normalization to its 1$\sigma$ uncertainty is $\approx 4.2$. The line is also evident in OBSID 2050260110, though at lower significance, with the ratio of the line normalization to its 1$\sigma$ uncertainty now at 3.5.  For OBSID 2050260109, the spectrum extracted closest to the outburst onset and at the lowest observed flux, we no longer find evidence for the line. When detected the line is narrow in the sense that it is unresolved and we can only place an upper limit on its width of $\approx 0.09$ keV (3$\sigma)$. Finally, in this work our primary focus is to model the X-ray spectrum to infer the broadband flux. Excluding the 1 keV line from the spectral fits only changes the inferred flux at the few percent level, so including it, or not, does not significantly alter our inferences regarding the source flux. We elected to include it since doing so provides a better overall statistical description of the data.

To estimate the outburst fluence we use simple linear interpolation between data gaps, and we also apply linear interpolation of the flux per unit count rates, based on the spectral results discussed above.  We employ the trapezoidal rule to integrate the total counts. We find a persistent emission energy fluence of $E_p = 7.92 \times 10^{-5}$ erg cm$^{-2}$, representing an estimate of the total energy associated with accretion from the outburst onset up to the initiation of the first observed burst. 

Assuming the observed, accretion-driven luminosity for spherical accretion,
\begin{equation}\label{eq:lp}
    L_{p} = 4\pi d^2 \xi_{p} f_p = \frac{z \dot M c^2}{(1+z)^3} \; ,
\end{equation}
where ${(1+z)=(1-2{GM}/{c}^{2}R)}^{-1/2}$, $\dot M$, $\xi_{p}$, $f_p$, $M$ and $R$ are the surface red-shift, mass accretion rate (as measured at the neutron star surface), persistent emission anisotropy factor, observed bolometric flux, neutron star mass and radius, respectively. We can use this equation to estimate the total accreted mass required to produce the observed energy fluence \citep{2020PhDT........10J}. We emphasize that $L_p$ and $f_p$ are the luminosity and flux as measured by an observer far from the neutron star.  The anisotropy factor, $\xi_p$, can be thought of as the solid angle into which the radiation is emitted, normalized by $4\pi$, thus, isotropic emission is characterized by $\xi_p = 1$.  We write the accreted mass column in the local neutron star frame as,
\begin{equation} 
y_a = \frac{M_a}{4\pi R^2} = \frac{\int \dot M(t')dt'}{4\pi R^2} \; ,
\end{equation}
where we use $t'$ to emphasize that the $\dot M$ integral is over the time as measured at the neutron star surface. With the use of equation (\ref{eq:lp}) this becomes,
\begin{equation} \label{eq:ya}
  y_a =  \frac{d^2 \xi_p (1+z)^2}{z c^2 R^2} \int f_p(t) (1+z) dt' = \frac{d^2 \xi_p (1+z)^2}{z c^2 R^2} E_p \; ,
\end{equation}
where $t$ is the time measured in the observer's frame, and we note that $dt = (1+z) dt'$, and thus $\int f_p(t)(1+z)dt' = \int f_p(t) dt = E_p$ is just the observed energy fluence.

If we assume $d = 3.5$ kpc \citep{2006ApJ...652..559G}, and use $E_p = 7.92 \times 10^{-5}$ erg cm$^{-2}$, we can write the accreted column as,
\begin{equation} \label{eq:ya2}
  y_a = 1.03 \times 10^7 \; \left ( \frac{\xi_p (1+z)^2}{z R_{10}^2} \right )\; \; \; {\rm g} \; {\rm cm}^{-2} \; , 
  \end{equation}
  where $R_{10}$ is the neutron star radius in units of 10 km. For $M = 1.4 M_{\odot}$, $R = 11$ km, and adopting $\xi_p = 1$ we find $y_a = 5.12 \times 10^7$ g cm$^{-2}$. With $M = 2.0 M_{\odot}$ and $R = 11$ km we find that $y_a$ decreases slightly to $3.91 \times 10^7$ g cm$^{-2}$.
  
  We can also use equation (\ref{eq:lp}) to estimate the mass accretion rate, $\dot M$, at the time of the weak burst onset. Using the estimated pre-burst flux of $7.9\times 10^{-10}$ erg cm$^{-2}$ s$^{-1}$, and a distance of $3.5$ kpc, we find,
  \begin{equation}
      \dot M = 2.03 \times 10^{-11} \; \left ( \frac{\xi_p (1 + z)^3}{z} \right )\; M_{\odot} \; {\rm yr}^{-1} \; .
  \end{equation}
  Using the same parameter assumptions as above, we find estimates of $\dot M = 1.56 \times 10^{-10}$ and $1.38 \times 10^{-10}$ $M_{\odot}$ yr$^{-1}$, respectively.  

\subsection{Burst Spectral Evolution: Peak Flux and Fluence}

\begin{figure}[htbp]
\includegraphics[width=\linewidth]{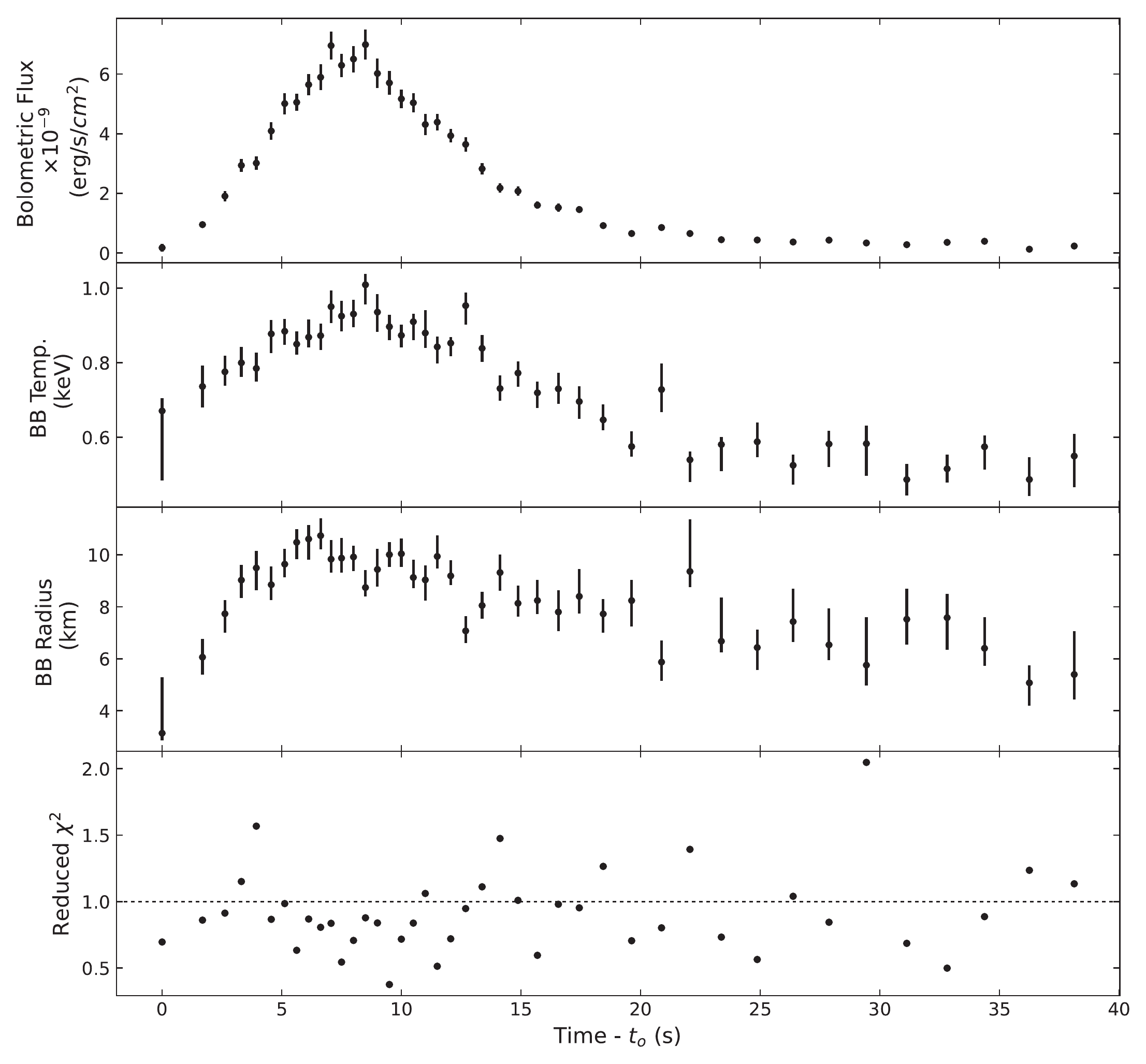}
\caption{\label{fig:evolution} Evolution of the weak X-ray burst derived from spectral modeling in the $0.5-10$ keV band. We show from the top down: the bolometric flux, blackbody temperature, blackbody radius (at 3.5 kpc), and reduced $\chi^2$, respectively. The error bars indicate 1-$\sigma$ confidence intervals.}
\end{figure}

We first segmented the burst light curve into intervals of approximately 500 counts using a 1/8 s time bin. We modeled the segmented spectra in the 0.5 - 10 keV band by adding a blackbody component to the pre-burst persistent emission model. The parameters of the persistent emission model were frozen to their best fit values, given in Table 1, only allowing the added blackbody component to vary, so that our model is
\texttt{phabs(constant(diskbb + bbodyrad + powerlaw + gaussian) + bbodyrad)}.
We first tried multiplying the persistent emission model by a constant \citep{2013ApJ...772...94W}, but found it was not statistically necessary, as it was possible to get a good fit with it left frozen at 1.0. The resulting evolution of the bolometric flux, the free parameters of the blackbody temperature and blackbody radius (at 3.5 kpc), along with the resulting $\chi^2$ are shown in Figure \ref{fig:evolution}. We found a peak bolometric burst flux of $f_b = 6.98 \pm 0.50 \times 10^{-9}$ erg s$^{-1}$ cm$^{-2}$. Using trapezoidal numerical integration of the flux, we calculated a bolometric fluence of $7.05 \pm 1.16 \times 10^{-8}$ erg cm$^{-2}$. The burst luminosity is defined as $L_b = 4 \pi d^2 \xi_b f_b$, where $\xi_b$ characterizes the anisotropy of the burst emission. Adopting $\xi_b =1$, and with $d = 3.5$ kpc, we can then estimate that the total energy released during the burst was $1.03 \times 10^{38}$ ergs.

\section{Physical Scenario and Interpretation}

Transient systems like J1808 provide an interesting laboratory to explore the different predicted regimes of nuclear burning on neutron stars.  Deep X-ray spectroscopic studies of the object in quiescence suggest rapid cooling of the neutron star core, perhaps by a form of enhanced neutrino emission such as Direct Urca \citep{2009ApJ...691.1035H}, which also provides some tentative evidence for a more massive neutron star ($\gtrsim 2 M_{\odot}$) in the system. 
Using the surface effective temperature constraints in quiescence from \cite{2009ApJ...691.1035H} and the theoretical results of \cite{1997A&A...323..415P}, \cite{2013ApJ...773..140M} estimated a core temperature for J1808 in the range from $7.2 - 7.7 \times 10^6$ K, for neutron star masses between 1.4 and 2.0 $M_{\odot}$.  Due to the high thermal conductivity in the core and crust \citep{2009ApJ...698.1020B}, it is thus very likely that when accretion begins in J1808 after a period of quiescence, the accumulating layer starts out at temperatures $\lesssim 1 \times 10^7$ K, that is, well below the temperature at which CNO cycle hydrogen-burning becomes thermally stable \citep{1981ApJ...247..267F, 2004NuPhS.132..435C, 2021ASSL..461..209G}. In this temperature-sensitive regime hydrogen-burning proceeds at very low levels, and the thermal profile of the accumulating layer will be set principally by compressional heating. This is a much less efficient heat source than the energy released from hot-CNO cycle burning in the stable burning regime.  

\cite{1981ApJ...247..267F} estimate the accretion rate required to maintain a stable hydrogen-burning shell (see their Table 1, $\dot M_{st}(B)$) as $2.7 \times 10^{-10}$ $M_{\odot}$ yr$^{-1}$, for a neutron star mass and radius of $1.41 M_{\odot}$ and 6.57 km. We note that the somewhat older neutron star models employed by \cite{1981ApJ...247..267F} have rather small radii compared to that suggested by more recent modeling \citep{2019ApJ...887L..24M, 2019ApJ...887L..21R}. For a more typical radius of, say, 11 km (which we employed above), we would expect that the estimated rate would increase modestly by $\approx 10 \%$, which would bring the value to $\approx 3 \times 10^{-10}$ $M_{\odot}$ yr$^{-1}$. Expressed as a fraction of the Eddington accretion rate, $\dot M_{Edd}$, and adopting the value of $\dot M_{Edd} = 1.8 \times 10^{-8}$ $M_{\odot}$ yr$^{-1}$ \citep{2004NuPhS.132..435C}, this is then equivalent to $\dot M = 0.0167 \; \dot M_{Edd}$. Note also that \cite{2004NuPhS.132..435C} quotes a value of $\dot M \gtrsim 0.01 \; \dot M_{Edd}$ for stable, hot-CNO cycle hydrogen-burning.  In addition, \cite{2007ApJ...661..468C} used a two-zone model to carry out a linear stability analysis to specifically explore the conditions under which hydrogen-triggered bursts can occur at low accretion rates, and found that they occur for rates $\lesssim 0.003 \; \dot M_{Edd}$. 

Above, we estimated an accretion rate at the time of the weak {\it NICER} burst in the range from $\approx 1.38 - 1.56 \times 10^{-10}$ $M_{\odot}$ yr$^{-1}$ ($\dot M \approx 0.0077 - 0.0087 \; \dot M_{Edd}$). This is less than the required rates estimated by \cite{2004NuPhS.132..435C} and \cite{1981ApJ...247..267F} for stable CNO burning, but slightly higher than the accretion rate obtained by \cite{2007ApJ...661..468C}.  These considerations provide strong evidence that in the initial outburst stage, accretion onto J1808 proceeds in an $\dot M$ range consistent with what \cite{1981ApJ...247..267F} refer to as {\it case 3} shell flashes. In this regime the accumulating layer remains cool enough that CNO hydrogen-burning proceeds in the temperature-sensitive regime,  that is, very little hydrogen is burned until the layer reaches the conditions for unstable ignition. Indeed, following \citet[see their equation 11]{1981ApJ...247..267F}, we would estimate that only about $ 1 - 2 \%$ of the hydrogen would be burned prior to ignition.  Further insights are provided by our estimates of the total column of matter accreted at the time of the burst, and its total energy fluence.  In \S 2 above we estimated the accreted column to be in the range $\approx 3.91 - 5.12 \times 10^{7}$ g cm$^{-2}$, and we measured an energy fluence in the burst of $\approx 1 \times 10^{38}$ erg (both at 3.5 kpc). For the following discussion we refer the reader to the illustrative hydrogen ignition curves presented by \citet[see their Fig. 1]{2004NuPhS.132..435C}, \citet[see their Fig. 2]{2021ASSL..461..209G}, and \citet[see their Fig. 4]{2007ApJ...654.1022P}.  Based on these curves, we can estimate that a column of this size would be ignited at a temperature in the range from $\approx 4 - 5 \times 10^7$ K. What happens upon ignition of the hydrogen?  The unstable burning will quickly heat the layer, raising the temperature to at least that at which the CNO energy generation rate saturates, but likely somewhat higher. \cite{1981ApJ...247..267F} estimate that only a small fraction, $\Delta X$, of the hydrogen needs to burn in order to raise the temperature. For a temperature change of $10^8$ K, they estimate  $\Delta X \approx 0.002$ (see their equation 12). 

After ignition of the hydrogen, two subsequent paths have been described in the literature.  First, if the ignition column is small enough, then an increase in its temperature may not cause it to cross the helium ignition curve, and additional accretion and/or an increase in the helium fraction is required before it will ignite. Alternatively, for deeper ignition columns, a temperature increase of a few $10^8$ K would render the shell unstable to helium ignition, promptly producing a mixed H/He burst.  We note that the work of \cite{2007ApJ...661..468C} and \cite{2007ApJ...654.1022P} also predict these two paths, and their calculations provide estimates of the hydrogen ignition columns that are broadly consistent with our estimate of the column accreted at the time of the weak burst. For example, at an accretion rate of $\dot m = 0.002 \; \dot m_{Edd}$ \citet[see their Fig. 4, right column]{2007ApJ...661..468C} find behavior consistent with the first scenario, a sequence of weak hydrogen flashes occurs until the helium column grows sufficiently to reach ignition conditions.  These calculations also provide an estimate of the temperature increase produced by the unstable hydrogen ignition, and suggest that changes of $\sim 2 \times 10^8$ K are likely. \citet[see their Fig. 7]{2007ApJ...654.1022P} also find a regime where hydrogen ignition does not lead to prompt  ignition of a He burst. They also explore the effect of sedimentation on hydrogen-triggered bursts, which enhances the amount of CNO nuclei at the ignition depth and causes a sharper temperature rise. Sedimentation is likely to play an important role in setting the ignition conditions for the weak burst given the low estimated accretion rate.

Measurement of the burst fluence enables us to estimate the fraction, $f_h$, of accreted hydrogen needed to burn in order to produce that much energy.  For an energy release (per gram) of $E_h = 6.4 \times 10^{18}$ erg g$^{-1}$ \citep{1983psen.book.....C}, we would require $m_h = 1.6 \times 10^{19} \; (1+z)$ g of hydrogen to burn, where the factor of $(1+z)$ is included because we are interested in the energy released at the neutron star surface. Expressed as a column on the neutron star, $y_h = m_h / 4\pi R^2$, and assuming $R = 11$ km, we find $y_h = 1.05 \times 10^{6} \; (1+z)$ g cm$^{-2}$. The amount of hydrogen present in the accreted column is $y_a X$, where $X$ is the mass fraction of hydrogen in the accreted material.
We thus have,
\begin{equation}
f_h = \frac{y_h}{y_a X} = 0.105 \; \frac{(1+z)}{Y_a X} \; ,
\end{equation} 
where $Y_a$ is the estimated accreted column in units of $10^7$ g cm$^{-2}$. Taking
$Y_a$ in the range from $3.9 - 5.1$, a fractional hydrogen abundance in the accreted fuel of $X = 0.7$, and using the same $M$ and $R$ assumptions employed above to evaluate $(1+z)$, we find $f_h$ in the range from $0.04 - 0.06$.  Note that this value should be considered a lower limit, as it assumes that the estimated total accreted column produced only a single such burst, and the mass fraction would likely be reduced further if sedimentation is present \citep[see below for additional discussion regarding potentially missed bursts]{2007ApJ...654.1022P}. This value is larger than the estimate given by \cite{1981ApJ...247..267F} for the fraction of hydrogen needed to burn to raise the temperature of the fuel by $\Delta T = 10^{8}$ K; we don't know the actual temperature increase however, and the estimate of \cite{1981ApJ...247..267F} should be thought of as a lower limit to our estimate from the measured burst energy fluence, since burning will continue at the stable CNO burning rate.  

Alternatively, we can ask the question, how much energy would we expect in the burst if the entire accreted column were to burn to heavy elements?  The energy release per gram, $Q_{nuc}$, would depend on the details of the nuclear burning pathways, however, employing the value of $Q_{nuc} = (1.3 + 5.8 X) \times 10^{18}$ erg g$^{-1}$ \citep{2006ApJ...652..559G}, and again adopting $X = 0.7$ we would expect $\approx 3.2 - 4.2 \times 10^{39}$ ergs liberated at the neutron star surface by burning all the fuel. This estimate also assumes that the total accreted column produces a single burst.  This is a factor of 30 $-$ 40 larger than the observed fluence in the weak X-ray burst, and also argues that the weak burst is likely not a mixed H/He burst. Rather, our analysis suggests that it likely represents the unstable ignition of a modest fraction of the hydrogen in the accreted layer, which constitutes strong observational evidence for such a weak ``hydrogen-only'' flash.

Interestingly, in their two zone model \cite{2007ApJ...661..468C} compute the peak energy fluxes produced during such weak hydrogen flashes. The range of fluxes that their model can produce is summarized in their Fig. 7 (bottom panel), where the peak flux is given as a fraction of the Eddington flux. Working backward, we measured a peak flux during the weak X-ray burst of $\approx 6.9 \times 10^{-9}$ erg cm$^{-2}$ s$^{-1}$.  If we scale this by the peak flux ($2.3 \times 10^{-7}$ erg cm$^{-2}$ s$^{-1}$) of the Eddington-limited burst observed later in the outburst \citep{2019ApJ...885L...1B}, we find a ratio of $\log (0.03) = -1.52$. Looking at Fig. 7 in \cite{2007ApJ...661..468C} (bottom panel), we can find hydrogen flashes (the dotted-dashed lines in the figure) in a narrow range of accretion rate that reach this flux level. 

\subsection{Stable burning after the burst?}

Previous theoretical studies concluded that the ignition of the hydrogen flash will raise the temperature in the layer to at least the stable burning regime, and likely higher.  Thus, hot-CNO cycle burning would be expected to continue for some period of time after the unstable ignition. Can we see evidence for such stable burning in the {\it NICER} data?  Interestingly, there is a clear ``offset'' between the pre- and post-burst flux levels.  This offset can be seen in Figure \ref{fig:lightcurburst}. Note that the inset panel uses a larger time bin size and log scale to emphasize the persistent count rate levels, to more clearly highlight the offset. We also plot the average count rate value for the pre-burst level (red dashed line) as a guide to the eye. To explore this question further we used the same spectral model to characterize the post-burst data as we used for the pre-burst and other persistent emission intervals. The time interval used for the post-burst spectral extraction is marked by the vertical dotted lines in Figure \ref{fig:lightcurburst} (main panel). We first tried to fit the post-burst spectrum using the same spectral shape as obtained from the pre-burst interval, allowing for the constant, $f_a$ parameter to make up the flux difference.  This did not provide an acceptable fit, and suggests the presence of an additional spectral component in the post-burst interval.  To explore this further we subtracted the pre-burst spectrum from the post-burst and found that the remaining excess could be well fit by a soft thermal spectrum, characterized as a blackbody with $kT = 0.51 \pm 0.02$ keV, normalization of $82.5 \pm 12.0$, and bolometric flux of $6.1 \pm 0.2 \times 10^{-11}$ erg cm$^{-2}$ s$^{-1}$. This is equivalent to a luminosity of $\approx 8.9 \times 10^{34} $ erg s$^{-1}$ (at 3.5 kpc).  

If the hydrogen burns stably at the same rate as it is accreted, then we would estimate a hydrogen-burning luminosity of $L_h = X\dot m E_h$, where $X$ is the mass fraction of hydrogen in the accreted fuel, $\dot m$ is the mass accretion rate at the burst onset, and $E_h$ is the energy production per gram due to hydrogen-burning.  With $\dot m = 1.4\times 10^{-10}$ $M_{\odot}$ yr$^{-1}$, $X=0.7$, and $E_h = 6.4\times 10^{18}$ erg g$^{-1}$, we would predict a stable hydrogen-burning luminosity of $\approx 4\times 10^{34}$ erg s$^{-1}$, which is a good fraction of the measured offset.  Perhaps a better estimate can be obtained by evaluating the energy production rate associated with the saturated, hot-CNO burning rate as, $L_{CNO} = 4\pi R^2 y_a \epsilon_{CNO}$, where $\epsilon_{CNO}$, $y_a$, and $R$, are the energy production rate due to hot-CNO burning, the accreted column depth, and the neutron star radius, respectively.  With $\epsilon_{CNO} = 5.8\times 10^{13} (Z_{CNO}/0.01)$ erg g$^{-1}$ s$^{-1}$ \citep{2000ApJ...544..453C}, $y_a = 4.5 \times 10^7$ g cm$^{-2}$, and $R = 11$ km, we find $L_{CNO} \approx 4\times 10^{34} (Z_{CNO}/0.01)$ erg s$^{-1}$. Here, $Z_{CNO}$ is the abundance of the CNO catalyzing elements. Employing the solar value $Z_{CNO} = 0.016$, we find $L_{CNO} = 6.4 \times 10^{34}$ erg s$^{-1}$, however, as noted above, at these low accretion rates sedimentation is very likely to be effective in enhancing the abundance of CNO elements near the base of the accreted fuel layer. For example, \citet[see their Figs. 2 \& 3]{2007ApJ...654.1022P} report enhancements in CNO element abundances by factors of 2 to 5, depending on the accretion rate.  

Based on these estimates it appears plausible that most or all of the observed flux offset can be accounted for by quasi-steady, hot-CNO burning of hydrogen. We note that the thermal nature of the spectral excess, and its $\approx 0.5$ keV temperature, similar to that at late times during the weak burst, is also consistent with this interpretation. This conclusion is also consistent with the hydrogen flash temperature and flux evolution calculations of \cite{2007ApJ...661..468C}. As an example, the hydrogen flashes shown in their Fig. 4 indicate that at ignition the flux rises abruptly, but then shows a ``plateau-like'' phase which decays over a timescale of several hours. The average flux levels near the beginning of these events are approximately consistent with the stable hydrogen-burning luminosity we estimated above. Once ignited, these flashes are burning hydrogen to helium in the fuel layer at essentially the saturated, hot-CNO cycle rate. We suggest that the two-zone model of \cite{2007ApJ...661..468C} (with H and He zones) probably does not adequately track and resolve the fast, initial hydrogen-burning when the thermal instability is initiated, but better predicts the longer timescale, thermally stable burning. The hydrogen-only ignition modeled by \citet[see their Fig. 7]{2007ApJ...654.1022P} also appears at least approximately similar to what is observed for the weak {\it NICER} burst. Indeed, 
the ratio of the peak burst bolometric flux to the
persistent, pre-burst flux is $7 \times 10^{-9} / 7.9 \times 10^{-10} \approx 8.8$, which is similar to the peak value of $F_{cool} / F_{acc}$ for the initial, burst-like flux increase shown in their Figure 7 (middle panel), and the overall burst duration appears consistent with the observed burst as well. 

More detailed radially resolved, and perhaps multidimensional calculations will likely be needed to more accurately track the rapid hydrogen ignition phase which we suggest may account for the weak {\it NICER} burst.  To briefly summarize, the weak {\it NICER} burst and post-burst flux offset appear to be consistent with the onset of a hydrogen-triggered shell flash in the cool, temperature-sensitive regime of the CNO cycle. The ignition column was likely shallow enough that the subsequent temperature increase was not sufficient to also promptly ignite a helium-burning instability.  

\subsection{Missed bursts?}

While {\it NICER} was able to begin observations quite close to the onset of accretion in the 2019 August outburst, the overall on-source coverage from onset to the time of the first observed burst was still rather modest, with a duty-cycle of about 4\%. Thus, if other bursts occurred it is conceivable that the {\it NICER} observations simply missed them. However, based on our estimate of the size of the accreted column, as well as current theoretical estimates of the hydrogen ignition curve, we argue that likely only a few such bursts might have been missed.  Firstly, while we don't know the precise temperature trajectory of the initial accumulating layer, it cannot plausibly be $\lesssim 2 \times 10^7$ K because at such low temperatures only columns much larger ($\gtrsim 2 - 3 \times 10^8$ gm cm$^{-2}$) than our estimate of the accreted column at the time of the weak burst ($3.9 - 5.1 \times 10^7$ g cm$^{-2}$) would be needed to ignite unstable burning, and such an ignition would also very likely lead to a bright, mixed H/He burst, which was not observed, though could have perhaps been missed. Secondly, as the temperature of the fuel layer increases the size of the unstable column decreases, however, above temperatures of about $8 \times 10^7$ K the hydrogen-burning will stabilize, precluding bursts.  This sets a minimum combustible column for hydrogen ignition which is, using the ignition curve in \cite{2004NuPhS.132..435C} as a guide, $\approx 1 \times 10^7$ g cm$^{-2}$.  Based on our estimated accreted column this would set a limit of not more than about five such bursts potentially being produced, as that would just about exhaust the total column accreted at the time of the weak burst. Another benchmark can be set by the accretion rate. We estimated a value of $\dot m = 1.4 - 1.6 \times 10^{-10}$ $M_{\odot}$ yr$^{-1}$ at the time of the weak burst.  If we take half of this value as more representative of the mean rate during the 72 hrs prior to the weak burst, we can estimate the time required to accrete the minimum unstable column of $1 \times 10^7$ g cm$^{-2}$. For $\dot m = 7\times 10^{-11}$ $M_{\odot}$ yr$^{-1}$, and assuming a radius $R = 11$ km, we find it would take 9.5 hr to accumulate such a column. Since the weak burst was observed after about 2.9 days, this also suggests an upper limit of $\sim 7$ to the total number of such weak bursts. We suggest that the actual temperature trajectory is probably somewhere between the two extremes described above, perhaps consistent with an unstable column on the order of $\sim 2-3 \times 10^7$ gm cm$^{-2}$. If this is correct it would suggest that the {\it NICER} observations may have missed one or two such weak bursts.  

\subsection{Other examples: RXTE observations of the 2005 outburst}

\begin{figure}[htbp]
\includegraphics[width=\linewidth]{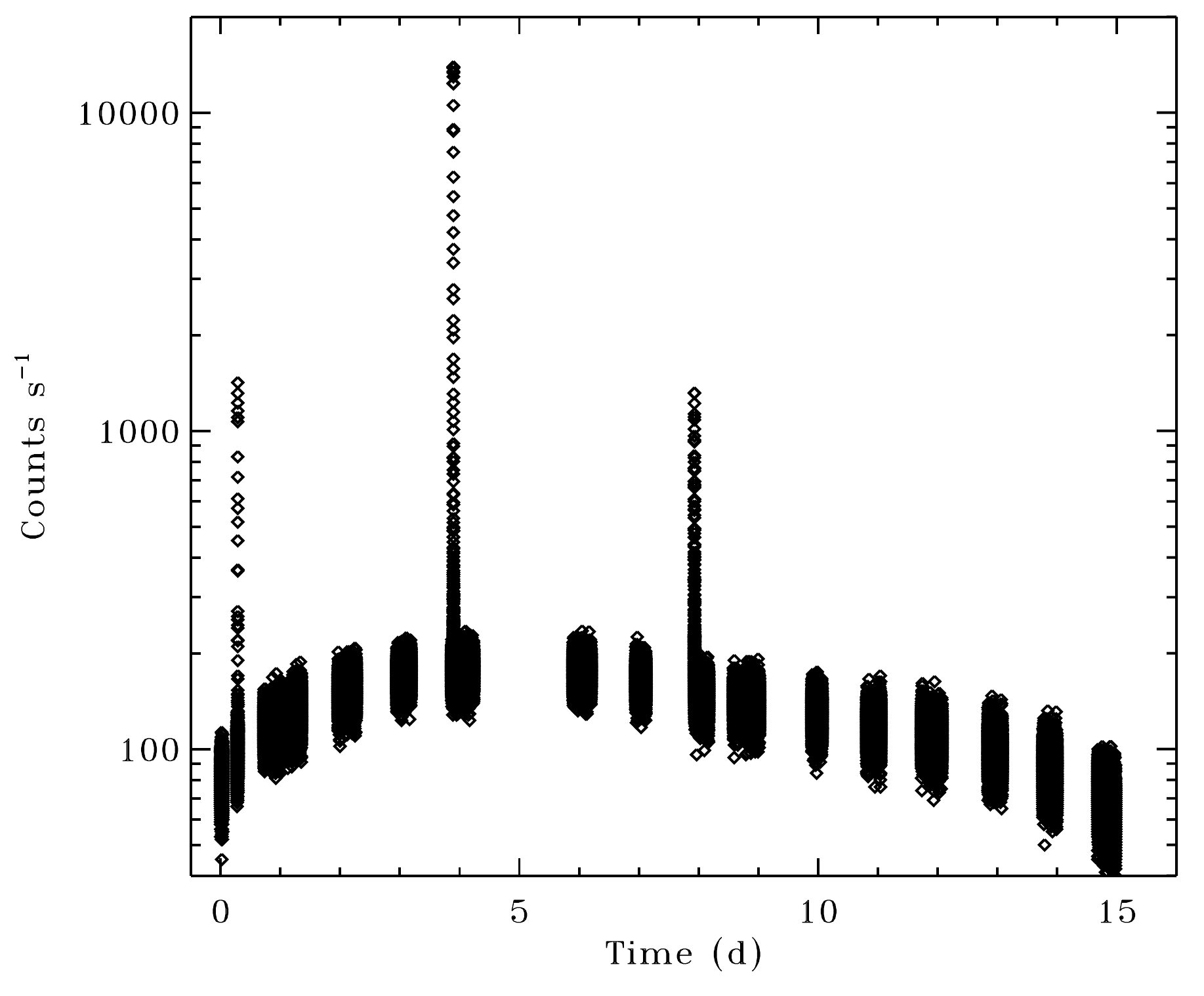}
\caption{\label{fig:rxteoutburst} Light curve from {\it RXTE} data (PCU 2, 3--30 keV) of the 2005 outburst from J1808. Note the logarithmic scale.  A weak X-ray burst is seen early in this outburst. Much brighter and energetic bursts are seen near days 4 and 8.  Note that the burst near day 8 was truncated by the RXTE exposure, and almost certainly the brightest part of this event was missed.  }
\end{figure}

We searched the literature and previous observations of J1808 to try and identify similar examples of weak bursts. We found a quite similar event early in the 2005 June outburst that was observed with {\it RXTE}.  We show in Figure~\ref{fig:rxteoutburst} the light curve of this outburst as obtained from {\it RXTE} pointed observations.  This burst occurred on 2 June at approximately 00:42:30 TT, and is evident near 0.25 days in the figure.  We carried out a time resolved spectral analysis of this event, and found qualitatively similar properties for this burst as for the weak {\it NICER} burst.  It reaches a peak bolometric flux of $1.54 \pm 0.11 \times 10^{-8}$ erg cm$^{-2}$ s$^{-1}$, about a factor of 2 greater than the {\it NICER} burst. It also had a peak blackbody temperature of 1.25 keV, which is about $25 \%$ larger than that of the {\it NICER} burst.  We note that this burst appears in the Multi-Instrument Burst Archive (MINBAR) catalog, with a reported peak bolometric flux of $1.6 \times 10^{-8}$ erg cm$^{-2}$ s$^{-1}$, and a fluence of $1.67 \times 10^{-7}$ erg cm$^{-2}$ \citep{2020ApJS..249...32G}.

The first evidence of active accretion for this outburst was provided by {\it RXTE/PCA} Galactic bulge scan observations on 31 May at 23:00:00 UTC, and indicated a persistent 2 - 10 keV X-ray flux level of $\approx 3$ mCrab \citep{2005ATel..505....1M}.  This flux value is similar to that measured with {\it NICER} for OBSID 2050260109 during the 2019 outburst (see Table~\ref{tab:persistent emission}).  The X-ray burst was observed approximately 25.7 hr later, and MINBAR reports a persistent flux at the time of the burst of $8.6 \times 10^{-10}$ erg cm$^{-2}$ s$^{-1}$, just a bit larger than the value estimated prior to the 2019 {\it NICER} burst (again, see Table~\ref{tab:persistent emission}).  We can use the pre-burst flux value reported by MINBAR and the earliest {\it RXTE} observations of the 2005 outburst reported by \cite{2005ATel..505....1M} and \cite{2005ATel..507....1W} to estimate the persistent, accretion-driven fluence prior to the weak 2005 burst. Evaluating a simple trapezoidal sum gives a value of $3.8 \times 10^{-5}$ erg cm$^{-2}$ that is approximately half of the estimated fluence prior to the 2019 {\it NICER} event.  This then suggests a total accreted column just prior to the 2005 {\it RXTE} event of about half that estimated for the 2019 {\it NICER} burst. Simply scaling our value estimated for the 2019 {\it NICER} burst suggests a range of $2.0 - 2.6 \times 10^{7}$ g cm$^{-2}$ for the total accreted column prior to the 2005 {\it RXTE} event.   

\subsection{Subsequent bursts detected with NuSTAR}

\begin{figure}[htbp]
\includegraphics[width=\linewidth]{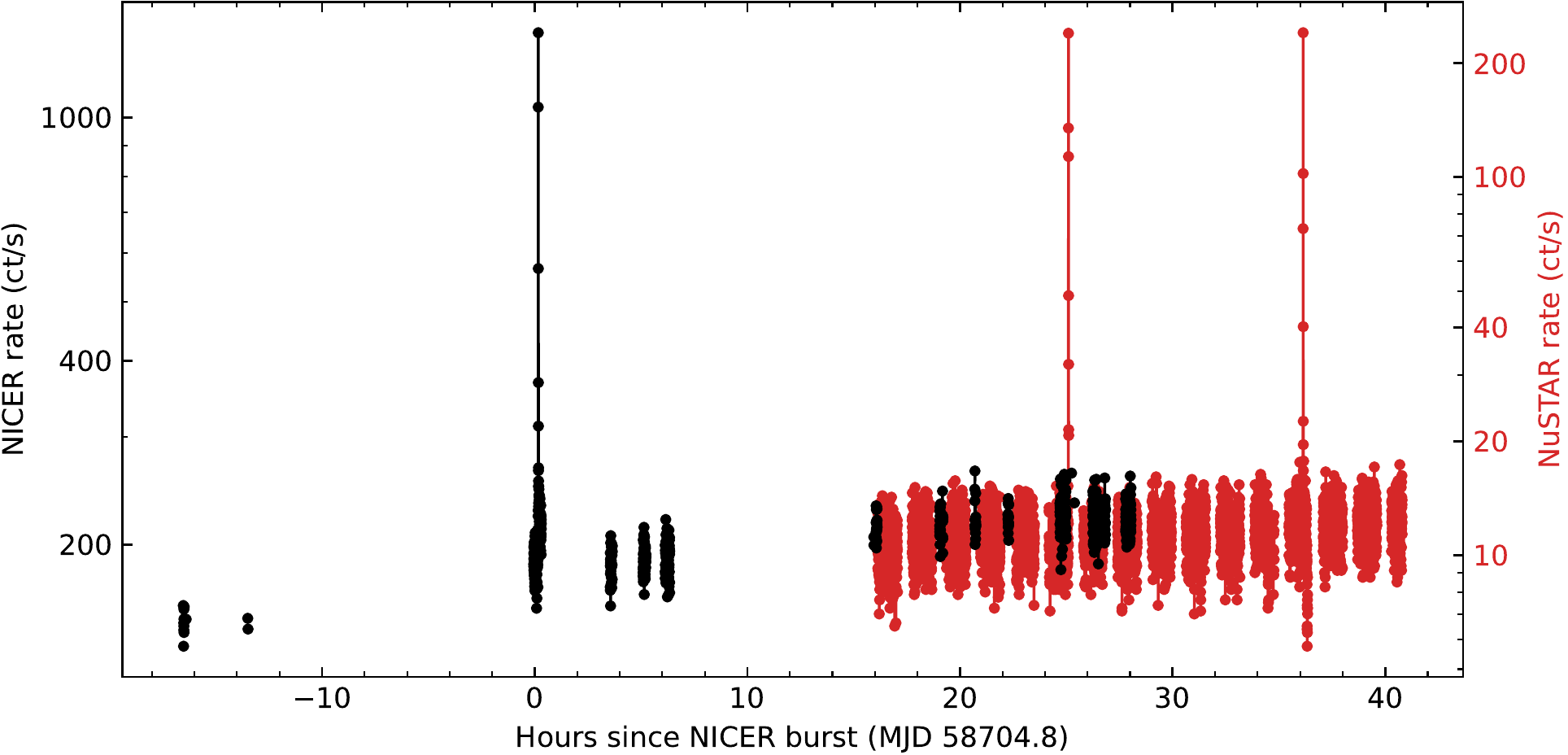}
\caption{Light curves from NICER (black, left axis) and NuSTAR (red, right axis) around
the time of the weak X-ray burst at t=0. Both light curves are calculated using an 8-s time resolution.
\label{fig:nicer nustar lc}}
\end{figure}

Additional observations of J1808 were collected with {\it NuSTAR} between 2019 August 10 and 11 (MJD 58705.5-58706.5). While these data do not cover the time of the weak X-ray burst observed with {\it NICER}, {\it NuSTAR} did catch two subsequent bursts, providing some additional, interesting context to this early phase of the outburst.
We processed the {\it NuSTAR} data (ObsID 90501335002) using \textsc{nustardas} version 2.1.2. Source data were extracted in the $3-79$ keV energy range from a $40\arcsec$ circular region centered on the source coordinates. The background was extracted using the same approach, but with the extraction region positioned in the background field.
The {\it NuSTAR} light curve reveals two X-ray bursts, the first of which occurred 24.8 hours after the weak {\it NICER} burst, while the second occurred another 11 hours later. This light curve is shown in Figure \ref{fig:nicer nustar lc}. We emphasize that though some of the NICER exposure was simultaneous with NuSTAR, this did not include these two bursts, and they were only observed with NuSTAR.

We first investigate the persistent emission by extracting a spectrum from a $100$ second window just prior to the first {\it NuSTAR} burst. As can be seen in Figure \ref{fig:nicer nustar lc}, this epoch was simultaneously observed with {\it NICER}, so we also extracted the contemporaneous {\it NICER} spectrum to obtain broadband energy coverage. We model this spectrum using the same persistent emission used previously (see Table~\ref{tab:persistent emission}), allowing for a constant cross calibration factor between {\it NICER} and FPMA/B of {\it NuSTAR}. In keeping with the analysis of the {\it NICER} burst, we extrapolated the best spectral model over $0.1-100$ keV to find a bolometric flux estimate of $1.47 \pm 0.05 \times 10^{-9}$ erg s$^{-1}$ cm$^{-2}$. 

From the recurrence times between the observed bursts, we obtain an estimate of the fluence due to the accretion of $1.3 \times 10^{-4}$ erg cm$^{-2}$ and $5.8 \times 10^{-5}$ erg cm$^{-2}$ for the two bursts, respectively. Converting these measurements to column depths, we use equation \ref{eq:ya2} to find $8.4 \times 10^7$ and $3.7 \times 10^7$ g cm$^{-2}$, respectively, where we again assumed a 1.4 $M_\odot$ neutron star mass and an $11$ km stellar radius. These column depths are of the same order as the one we calculated for the initial {\it NICER} burst.  Indeed, given the observed 11 hr recurrence time between the two {\it NuSTAR} bursts, and the relatively constant persistent flux (and hence accretion rate), it is conceivable that a similar burst was missed in the gap between the weak {\it NICER} burst and the first {\it NuSTAR} burst.  If so, then the accretion fluence for the two {\it NuSTAR} bursts would be essentially consistent with each other.  

To explore the burst spectra, we proceeded by dividing the bursts into multiples of 1/8 seconds such that each bin contains at least 500 counts. We extract a spectrum for each bin and model it using an absorbed blackbody in addition to the fixed persistent emission. The inferred burst properties obtained from these fits are listed in Table \ref{tab:burst properties}.
The two {\it NuSTAR} bursts had fluences of $5\times 10^{-7}$ erg cm$^{-2}$ and $3 \times 10^{-7}$ erg cm$^{-2}$, respectively. This means that they are about a factor of 4 - 7 times more energetic than the weak X-ray burst observed with {\it NICER}. The first {\it NuSTAR} burst reached a peak flux ten times greater than that of the weak {\it NICER} burst, and it was also significantly ``hotter,'' reaching a peak blackbody temperature of $2.3$ keV.  At the same time, these bursts remain much fainter than the Eddington-limited bursts observed at later times in the outbursts of J1808, which typically have fluences of $2\sim4\times 10^{-6}$ erg cm$^{-2}$ \citep{2008ApJS..179..360G, 2013A&A...553A..83I}.

\begin{table*}[t]
\caption{Burst parameters}
\label{tab:burst properties}
\begin{center}
\begin{tabular}{ccccc}
\tableline\tableline
Parameter & NICER 1 & NuSTAR 1 & NuSTAR 2 & NICER 2\\
\tableline\tableline
Onset (MJD, TT) & 58704.8068 & 58705.8459 & 58706.3058 & 58716.0861 \\
Peak flux (erg s$^{-1}$ cm$^{-2}$) & $7 \times 10^{-9}$ & $7 \times 10^{-8}$ & $4 \times 10^{-8}$  &  $3 \times 10^{-7}$ \\
Burst fluence (erg cm$^{-2}$) & $7 \times 10^{-8}$ & $5 \times 10^{-7}$ & $3 \times 10^{-7}$ & $2 \times 10^{-6}$\\
Accretion fluence (erg cm$^{-2}$) & $8 \times 10^{-5}$ & $1.3 \times 10^{-4}$ & $5.8 \times 10^{-5}$ & \nodata \\
Peak kT (keV) & 1.0 & 2.3 & 1.7 & 2.5 \\
\tableline\tableline
\end{tabular}
\end{center}
\tablecomments{The properties of the second NICER burst are
taken from \citet{2019ApJ...885L...1B} as an example of a
bright Eddington-limited X-ray burst from J1808}
\end{table*}

\section{Summary, Caveats \& Outlook}

Based on the considerations above we suggest a scenario similar to that discussed in the work of \cite{2007ApJ...661..468C} and \cite{2007ApJ...654.1022P} as a working hypothesis to account for the weak bursts observed by {\it NICER} and {\it RXTE} during the early days of the 2019 and 2005 outbursts of J1808.  As accretion begins, the neutron star is cool enough and the accretion rate is low enough that CNO hydrogen-burning in the accumulating layer occurs in the temperature-sensitive regime. At these lower temperatures, $\lesssim 5 \times 10^7$ K, very little hydrogen is burned. Significant burning of hydrogen will only begin when the accumulated column reaches the conditions for the thermal instability to set in.  For a temperature of $\approx 5 \times 10^7$ K this will occur at a column depth of about $3 \times 10^7$ g cm$^{-2}$. This value is not too dissimilar from the column estimated just prior to the 2005 event.  When the initial accumulating layer reaches ignition depth the hydrogen instability occurs, triggering a hydrogen flash.  We suggest that the initial rapid increase in the nuclear energy generation rate ultimately results in the ``heat pulse'' that is observed as the weak X-ray burst, however, we think that more sophisticated, multi-dimensional theoretical calculations of the time-dependent nuclear energy generation coupled with the subsequent heat and radiation transport, will be needed to test the details of this hypothesis.  After the initial hydrogen ignition, the burning layer will reach a high enough temperature that subsequent hydrogen-burning can proceed at the thermally stable level appropriate to the hot-CNO cycle. Above, we have argued that the observed offset between the pre- and post-burst flux levels of the 2019 event is consistent with this ``quasi-steady'' burning phase.  This source of heat will keep the layer warm enough for burning to continue for a time, likely measured in hours if conditions are not too dissimilar from those modeled by \cite{2007ApJ...661..468C}. During this time the quasi-stable burning will increase the helium fraction of the layer.  Given the gaps in {\it NICER} coverage after the weak burst, we cannot say how long this ``quasi-steady'' burning may have persisted, but we note that observations $\approx 3.5$ hrs after the burst show a count rate and flux approximately consistent with the pre-burst level.   For the conditions described above, that is, a hydrogen ignition column of $\approx 3\times 10^7$ g cm$^{-2}$, such an initial hydrogen flash is unlikely to produce a prompt helium ignition, simply because at that column depth the helium will not be thermally unstable \citep{2004NuPhS.132..435C}.  

As accretion continues, the hydrogen layer or layers that initially flashed will be pushed deeper, to higher column depths.  The freshly accreted material above it will also reach the hydrogen ignition depth, and if so, produce another hydrogen flash, assuming its temperature is low enough. In this way, a sequence of hydrogen flashes could be produced.  Eventually, the helium-enriched layers will likely reach column depths where the helium will ignite, producing  more energetic, mixed H/He bursts.  We suggest that the observed {\it NuSTAR} bursts are the result of this process. The steadily increasing accretion rate will also be an important variable, as this will tend to increase the temperature of the accreting layers.  More complete theoretical modeling of this process will have to include the time-varying accretion rate \citep{2018MNRAS.477.2112J}. 

If the above scenario is approximately correct we can try to speculate further regarding a few other details of the observations.  The 2005 event observed with {\it RXTE} was the earlier event in terms of the time since outburst onset, occurring approximately 1 day after onset.  Other things being equal one would expect the accreting layer to be cooler than at later times, such as the 2.9 days post-outburst from the 2019 event.  A cooler shell will have a larger unstable column, so that this could perhaps explain the fact that the {\it RXTE} event is the more energetic of the two weak bursts.  This also provides some tentative evidence that the 2019 {\it NICER} event may have been preceded by at least one additional weak burst that was missed.  

\subsection{Remaining Uncertainties and Alternative Interpretations}

In estimating accretion rates and accreted columns we allowed for variation in the neutron star mass, however, there are other uncertainties which complicate such estimates.  These include the source distance, anisotropy factors, bolometric corrections, and the line-of-sight absorption. We note that the more recent work of \cite{2019MNRAS.490.2228G} reports a slightly closer distance of $3.3^{+0.3}_{-0.2}$ kpc for J1808.   While their quoted uncertainty range includes the 3.5 kpc value we have adopted, a decrease from 3.5 to 3.3 kpc would reduce our estimates by a factor of $0.9$. These authors also provide estimates of the anisotropy factors for both persistent and burst emission, finding $\xi_p = 0.87^{+0.12}_{-0.10}$ and $\xi_b = 0.74^{+0.10}_{-0.10}$. Applying these values would also reduce the estimated accretion rate and column, by a factor of 0.87, and decrease our estimate of the burst peak luminosity and fluence, by a factor of 0.74. Adopting the best values reported by \cite{2019MNRAS.490.2228G} for both $d$ and $\xi_p$ would reduce the estimated accretion rate and accreted column by a factor of 0.77.

We have argued that the weak bursts result from, principally, hydrogen-burning, but are there other possibilities involving the unstable burning of helium? One scenario that can produce weak or underluminous bursts is the phenomenon of short recurrence time (SRT) bursts \citep{2010ApJ...718..292K}. The idea behind SRT events is that they burn fuel remaining from a preceding burst. In the present case, a preceding, larger burst would have had to occur (and been missed) for this idea to be workable.  In principle, this could account for the observed weak bursts, but there are some difficulties with this interpretation.  First, the estimated accreted columns are uncomfortably low. This scenario would require that a relatively bright, mixed H/He burst would have occurred prior to the observed weak events, and been missed in each case.  As discussed above, this would require relatively large ignition columns, likely $\gtrsim 2 \times 10^8$ g cm$^{-2}$, which is much larger than the estimated columns present just prior to each weak event. Our accretion column estimates would have to be underestimated by factors of 4 - 5 for this to be more plausible. Second, J1808 is not currently known to produce SRT events. There has been reasonably good coverage of past J1808 outbursts, and no SRT events have been definitively observed. For example, the compilation of SRT burst observations by  \cite{2010ApJ...718..292K} does not include J1808, and we also note that the 401 Hz spin frequency for J1808 is less than the faster, $\gtrsim 500$ Hz, spins associated with some of the known SRT sources. Third, the flux offset between the pre- and post-burst emission seems to make more sense in the context of stable hydrogen-burning than what might be expected from an SRT event, for which one would not typically expect to find such a flux offset. We note also that the theoretical mechanism of opacity-driven convective mixing explored by \cite{2017ApJ...842..113K} to account for SRT bursts occurs for ignition in relatively hot envelopes, which seems less applicable to the low accretion rate regime near burst onset that we have described above.  It is difficult to completely rule out the SRT scenario, but we think the considerations above argue against it. 

We have argued that the early accretion outburst evolution onto a ``cool'' neutron star in J1808 provides a unique environment to explore the physics of nuclear burning on neutron stars, and most interestingly, the ignition of unstable hydrogen-burning in the temperature-sensitive regime of the CNO cycle.  We suggest that the weak bursts seen by {\it NICER} and {\it RXTE} in the 2019 and 2005 outbursts, respectively, may result from this process.  More complete, continuous observational coverage of the first 4-5 days of subsequent outbursts from J1808 could definitively test this hypothesis.  Such data would also provide for detailed physical comparisons with new theoretical efforts to track the outcome of time-varying accretion onto neutron stars and the subsequent nuclear burning of the accreted matter.  This could provide interesting constraints on such things as the accretion rate, the thermal profile of the accreting matter and the nuclear energy generation and subsequent heat transport in the accreted layers.  

\begin{acknowledgments}
This work was supported by NASA through the {\it NICER} mission and the Astrophysics Explorers Program. This research also made use of data and/or software provided by the High Energy Astrophysics Science Archive Research Center (HEASARC), which is a service of the Astrophysics Science Division at NASA/GSFC and the High Energy Astrophysics Division of the Smithsonian Astrophysical Observatory. We thank Ed Brown, Andrew Cumming, and Duncan Galloway for comments that helped to improve this manuscript. P.B. acknowledges support from NASA through the Astrophysics Data Analysis Program (80NSSC20K0288) and the CRESST II cooperative agreement (80GSFC21M0002).
\end{acknowledgments}

\facility{NICER, ADS, HEASARC}

\software{NICERDAS (v8), XSPEC (Arnaud 1996)}

\bibliographystyle{aasjournal}

\bibliography{ms}

\end{document}